\documentclass[a4paper,12pt]{article}

\usepackage{fullpage}
\usepackage{authblk}
\usepackage{amsmath}
\usepackage{gensymb}
\usepackage{xfrac}
\usepackage{url}
\usepackage{csquotes}
\usepackage{graphicx}
\usepackage{enumitem}
\usepackage{multirow}
\usepackage[utf8]{inputenc}
\usepackage[bottom,hang,flushmargin]{footmisc}

\MakeOuterQuote{"}
\setlist{nolistsep}
\setlist[itemize]{leftmargin=*}
\setlist[enumerate]{leftmargin=*}

\title{cvBMS and cvBMA: filling in the gaps}
\author{Joram Soch}
\affil{BCCN Berlin, Germany}
\affil{joram.soch@bccn-berlin.de}
\date{}

\begin{document}

\setcounter{page}{0}
\maketitle
\vspace{-1em}

\small
\setlength{\leftskip}{2em}
\setlength{\rightskip}{2em}

\vspace{-0.5em} \begin{center} \textbf{Abstract} \end{center} \vspace{-0.5em}
With this technical report, we provide mathematical and implementational details of cross-validated Bayesian model selection (cvBMS) and averaging (cvBMA) that could not be communicated in the corresponding peer-reviewed journal articles. This will allow statisticians and developers to comprehend internal functionalities of cvBMS and cvBMA for further development of these techniques.

\setlength{\leftskip}{0pt}
\setlength{\rightskip}{0pt}
\normalsize

\small
\setlength{\leftskip}{2em}
\setlength{\rightskip}{2em}

\vspace{+0.5em} \begin{center} \textbf{Keywords} \end{center} \vspace{-0.5em}
fMRI-based neuroimaging, mass-univariate GLM, model selection, model averaging, cross-validation, Bayesian statistics

\setlength{\leftskip}{0pt}
\setlength{\rightskip}{0pt}
\normalsize

\vspace{1em}
\tableofcontents

\pagebreak
\section{Introduction} \label{sec:Introduction}

With \textit{cross-validated Bayesian model selection} (cvBMS; Soch et al., 2016) and \textit{cross-validated Bayesian model averaging} (cvBMA; Soch et al., 2017), we have recently described novel methods to achieve model quality control for general linear models (GLMs) applied to functional magnetic resonance imaging (fMRI) data (Soch, 2017). These methods were also published as an SPM toolbox for \textit{model assessment, comparison and selection} (MACS; Soch \& Allefeld, 2018).

For several reasons, the description of such methods in the peer-reviewed literature always lacks some details that are uninteresting to the average user, but can be crucial to know for the advanced developer. In this note -- which also serves as a meta paper to previous publications on these methods --, we review those details which are required to understand the mathematical structure of cvBMS and cvBMA. We hope that this will stimulate further development of these techniques.

When a \verb|function_name| appears in typewriter front, this refers to a routine of the MACS toolbox for SPM (DOI: 10.5281/zenodo.845404) which is freely available from GitHub (URL: \url{https://github.com/JoramSoch/MACS}). The reader of this text should be familiar with the cvBMS and cvBMA papers.

\section{Cross-validated Bayesian model selection (cvBMS)} \label{sec:cvBMS}

\subsection{Efficient calculation of the cross-validated LME}

The \textit{cross-validated log model evidence} (cvLME) is at the heart of cvBMS (and cvBMA). It is defined as

\begin{equation} \label{eq:cvLME}
\mathrm{cvLME}(m) = \sum_{i=1}^{S} \log \int p(y_i|\theta,m) \, p(\theta|\cup_{j \neq i} \, y_j,m) \, \mathrm{d}\theta
\end{equation}

where $m$ is a \textit{general linear model} (GLM) and $S$ is the number of fMRI recoding sessions or \textit{cross-validation} (CV) \textit{folds}. In a given CV fold $i$, $\cup_{j \neq i} \, y_j$ are all data except those from session $i$ (the "training data") and $y_i$ are the data from session $i$ (the "test data"). When an fMRI data set does not have mutliple sessions (\verb|MA_cvLME_multi|), the data from a single session are separated in two parts -- discarding between 10 and 19 scans in the middle to ensure temporal independence -- and split-half cross-validation with $S = 2$ is performed (\verb|MA_cvLME_single|).

One addend of the cvLME is referred to as an \textit{out-of-sample log model evidence} (oosLME). It is defined as

\begin{equation} \label{eq:oosLME}
\mathrm{oosLME}_i(m) = \log \int p(y_i|\theta,m) \, p(\theta|\cup_{j \neq i} \, y_j,m) \, \mathrm{d}\theta
\end{equation}

which essentially is a form of a \textit{log model evidence} (LME)

\begin{equation} \label{eq:LME}
\mathrm{LME}(m) = \log p(y|m) = \log \int p(y|\theta,m) \, p(\theta|m) \, \mathrm{d}\theta
\end{equation}

where $\theta$ are parameters of the model $m$, i.e. $\theta = \left\lbrace \beta, \tau \right\rbrace$ for a GLM with regression coefficients $\beta$ and residual precision $\tau$ (Soch et al., 2016, eq.~3). Whereas the LME uses a distribution $p(\theta|m)$ specified \textit{a priori}, the oosLME uses the distribution $p(\theta|\cup_{j \neq i} \, y_j,m)$ obtained from independent training data.

Calculating the oosLME for a given CV fold $i$ therefore proceeds in three steps: First, a posterior distribution is derived from the training data and a non-informative prior:

\begin{equation} \label{eq:train-post}
p(\theta|\cup_{j \neq i} \, y_j) \propto p(\cup_{j \neq i} \, y_j|\theta) \, p(\theta) \; .
\end{equation}

Second, another posterior distribution is derived from the test data and the posterior obtained from the training data which is now serving as a prior distribution:

\begin{equation} \label{eq:test-post}
p(\theta|\cup_{j \neq i} \, y_j, y_i) \propto p(y_i|\theta) \, p(\theta|\cup_{j \neq i} \, y_j) \; .
\end{equation}

Thus, the posterior obtained from the test data is a function of all the data and identical to the posterior that would be obtained from all data with the non-informative prior:

\begin{equation} \label{eq:all-post}
p(\theta|\cup_{j \neq i} \, y_j, y_i) = p(\theta|y) \propto p(y|\theta) \, p(\theta) \; .
\end{equation}

Also, note that the LME can always be written as a function of the data $y$, parameters of the prior distribution $p(\theta|m)$ and parameters of the posterior distribution $p(\theta|y,m)$ (Soch et al., 2016, eq.~9). Thus, the oosLME in the $i$-th CV fold is only a function of the test data $y_i$, hyperparameters of the posterior distribution from the training data $p(\theta|\cup_{j \neq i} \, y_j)$ -- now used as a prior distribution for the test data -- and hyperparameters of the posterior distribution from the test data $p(\theta|\cup_{j \neq i} \, y_j, y_i)$.

Because the posterior distribution from the test data is the same for all CV folds $i$ (see eq.~\ref{eq:all-post}), it only needs to be calculated once. This is, among other means such as sparingly-used loops and vector-based computations in \verb|ME_GLM_NG| and \verb|ME_GLM_NG_LME|, one element in an efficient calculation of the cvLME for the GLM. As the non-informative prior distrubtion, we use a flat Gaussian on the regression coefficients $\beta$ and Jeffrey's prior for the residual precision $\tau$ (Soch et al., 2016, eqs.~4/15):

\begin{equation} \label{eq:GLM-NG-prior-ni}
\begin{split}
p(\beta|\tau) &= \mathrm{N}(\beta; \mu_0, (\tau \Lambda_0)^{-1}) \\
p(\tau) &= \mathrm{Gam}(\tau; a_0, b_0) \\
\mu_0 &= 0_{p}, \; \Lambda_0 = 0_{pp} \\
a_0 &= 0, \; b_0 = 0 \; .
\end{split}
\end{equation}

Formulas for posterior hyperparameters and log model evidence of this \textit{general linear model with normal-gamma priors} (GLM-NG) are given in the cvBMS paper (Soch et al., 2016, eqs.~6/9) and implemented in \verb|ME_GLM_NG| and \verb|ME_GLM_NG_LME| which are being called from \verb|MA_cvLME_multi| and \verb|MA_cvLME_single| (Soch \& Allefeld, 2018, fig.~1).

\pagebreak
\subsection{Detailed derivation of accuracy and complexity for the GLM}

As outlined in the cvBMS paper, the LME (and the cvLME) can be partitioned into \textit{model accuracy} and \textit{model complexity} where the accuracy term is a \textit{posterior expected log-likelihood} (PLL) and the complexity penalty is a \textit{Kullback-Leibler} (KL) \textit{divergence} between posterior and prior distribution (Soch et al., 2016, eq.~12):

\begin{equation} \label{eq:LME-AnC}
\begin{split}
\mathrm{LME}(m) &= \mathrm{Acc}(m) - \mathrm{Com}(m) \\
\mathrm{Acc}(m) &= \left\langle \log p(y|\theta,m) \right\rangle_{p(\theta|y,m)} \\
\mathrm{Com}(m) &= \mathrm{KL} \left[ p(\theta|y,m) || p(\theta|m) \right] \; .
\end{split}
\end{equation}

In what follows, we will assume the GLM-NG with the likelihood function

\begin{equation} \label{eq:GLM-NG-LF}
p(y|\beta,\tau) = \mathrm{N}(y; X \beta, (\tau P)^{-1}) = \sqrt{\frac{|\tau P|}{(2 \pi)^n}} \, \exp\left[ -\frac{\tau}{2} (y-X\beta)^T P (y-X\beta) \right] \; ,
\end{equation}

the prior distribution

\begin{equation} \label{eq:GLM-NG-prior}
p(\beta,\tau) = p(\beta|\tau) \, p(\tau) = \mathrm{N}(\beta; \mu_0, (\tau \Lambda_0)^{-1}) \, \mathrm{Gam}(\tau; a_0, b_0)
\end{equation}

and the posterior distribution

\begin{equation} \label{eq:GLM-NG-post}
p(\beta,\tau|y) = p(\beta|\tau,y) \, p(\tau|y) = \mathrm{N}(\beta; \mu_n, (\tau \Lambda_n)^{-1}) \, \mathrm{Gam}(\tau; a_n, b_n) \; .
\end{equation}

\subsubsection{Model accuracy for the GLM-NG}

For the GLM-NG, the accuracy term amounts to

\begin{equation} \label{eq:GLM-NG-Acc1}
\begin{split}
\mathrm{Acc}(m) &= \iint p(\beta,\tau|y) \, \log p(y|\beta,\tau) \, \mathrm{d}\beta \, \mathrm{d}\tau \\
&= \int p(\tau|y) \int p(\beta|\tau,y) \, \log p(y|\beta,\tau) \, \mathrm{d}\beta \, \mathrm{d}\tau \\
&= \left\langle \left\langle \log p(y|\beta,\tau) \right\rangle_{p(\beta|\tau,y)} \right\rangle_{p(\tau|y)} \; .
\end{split}
\end{equation}

Plugging in the log-likelihood function gives

\begin{equation} \label{eq:GLM-NG-Acc2}
\begin{split}
\mathrm{Acc}(m) &= \left\langle \left\langle \frac{1}{2} \log |P| + \frac{n}{2} \log \tau - \frac{n}{2} \log (2 \pi) - \frac{1}{2} (y-X\beta)^T (\tau P) (y-X\beta) \right\rangle_{p(\beta|\tau,y)} \right\rangle_{p(\tau|y)} \\
&= \left\langle \left\langle \frac{1}{2} \log |P| + \frac{n}{2} \log \tau - \frac{n}{2} \log (2 \pi) - \frac{\tau}{2} \left[ y^T P y - 2 y^T P X \beta + \beta^T X^T P X \beta \right] \right\rangle_{p(\beta|\tau,y)} \right\rangle_{p(\tau|y)}
\end{split}
\end{equation}

\pagebreak
If $x \sim \mathrm{N}(\mu, \Sigma)$, then

\begin{equation} \label{eq:mvn-exp}
\begin{split}
\left\langle x \right\rangle &= \mu \\
\left\langle x^T A x \right\rangle &= \mu^T A \mu + \mathrm{tr}(A \Sigma)
\end{split}
\end{equation}

from which follows that

\begin{equation} \label{eq:GLM-NG-Acc3}
\begin{split}
\mathrm{Acc}(m) &= \left\langle \frac{1}{2} \log |P| + \frac{n}{2} \log \tau - \frac{n}{2} \log (2 \pi) - \right. \\
&\hphantom{= -} \left. \frac{\tau}{2} \left[ y^T P y - 2 y^T P X \mu_n + \mu_n^T X^T P X \mu_n + \frac{1}{\tau} \mathrm{tr}(X^T P X \Lambda_n^{-1}) \right] \right\rangle_{p(\tau|y)} \; .
\end{split}
\end{equation}

If $x \sim \mathrm{Gam}(a, b)$, then

\begin{equation} \label{eq:Gam-exp}
\begin{split}
\left\langle x \right\rangle &= \frac{a}{b} \\
\left\langle \log x \right\rangle &= \psi(a) - \log(b)
\end{split}
\end{equation}

which gives the final result (Soch et al., 2016, eq.~C.2):

\begin{equation} \label{eq:GLM-NG-Acc4}
\begin{split}
\mathrm{Acc}(m) = & - \frac{1}{2} \frac{a_n}{b_n} (y-X\mu_n)^T P (y-X\mu_n) - \frac{1}{2} \mathrm{tr}(X^T P X \Lambda_n^{-1}) \\
& + \frac{1}{2} \log |P| - \frac{n}{2} \log (2 \pi) + \frac{n}{2} (\psi(a_n) - \log(b_n))  \; .
\end{split}
\end{equation}

\subsubsection{Model complexity for the GLM-NG}

For the GLM-NG, the complexity penalty amounts to

\begin{equation} \label{eq:GLM-NG-Com1}
\begin{split}
\mathrm{Com}(m) &= \iint p(\beta,\tau|y) \, \log \frac{p(\beta,\tau|y)}{p(\beta,\tau)} \, \mathrm{d}\beta \, \mathrm{d}\tau \\
&= \iint p(\beta|\tau,y) \, p(\tau|y) \, \log \left[ \frac{p(\beta|\tau,y)}{p(\beta|\tau)} \, \frac{p(\tau|y)}{p(\tau)} \right] \, \mathrm{d}\beta \, \mathrm{d}\tau \\
&= \int p(\tau|y) \int p(\beta|\tau,y) \, \log \frac{p(\beta|\tau,y)}{p(\beta|\tau)} \, \mathrm{d}\beta \, \mathrm{d}\tau + \int p(\tau|y) \, \log \frac{p(\tau|y)}{p(\tau)} \int p(\beta|\tau,y) \, \mathrm{d}\beta \, \mathrm{d}\tau \\
&= \left\langle \mathrm{KL} \left[ p(\beta|\tau,y) || p(\beta|\tau) \right] \right\rangle_{p(\tau|y)} + \mathrm{KL} \left[ p(\tau|y) || p(\tau) \right] \; .
\end{split}
\end{equation}

With the multivariate normal KL divergence (Soch \& Allefeld, 2016a, eq.~8)

\begin{equation} \label{eq:mvn-KL}
\mathrm{KL} \left[ \mathrm{N}(x; \mu_1, \Sigma_1) || \mathrm{N}(x; \mu_2, \Sigma_2) \right] = \frac{1}{2} \left[ (\mu_2 - \mu_1)^T \Sigma_2^{-1} (\mu_2 - \mu_1) + \mathrm{tr}(\Sigma_2^{-1} \Sigma_1) - \log \frac{|\Sigma_1|}{|\Sigma_2|} - k \right]
\end{equation}

\pagebreak
and the univariate Gamma KL divergence (Soch \& Allefeld, 2016a, eq.~12)

\begin{equation} \label{eq:Gam-KL}
\mathrm{KL} \left[ \mathrm{Gam}(x; a_1, b_1) || \mathrm{Gam}(x; a_2, b_2) \right] = a_2 \, \log \frac{b_1}{b_2} - \log \frac{\Gamma(a_1)}{\Gamma(a_2)} + (a_1 - a_2) \, \psi(a_1) - (b_1 - b_2) \, \frac{a_1}{b_1} \; ,
\end{equation}

we obtain the model complexity as

\begin{equation} \label{eq:GLM-NG-Com2}
\begin{split}
\mathrm{Com}(m) &= \left\langle \frac{1}{2} \left[ (\mu_0 - \mu_n)^T (\tau \Lambda_0) (\mu_0 - \mu_n) + \mathrm{tr}((\tau \Lambda_0) (\tau \Lambda_n)^{-1}) - \log \frac{|(\tau \Lambda_n)^{-1}|}{|(\tau \Lambda_0)^{-1}|} - p \right] \right\rangle_{p(\tau|y)} \\
&+ a_0 \, \log \frac{b_n}{b_0} - \log \frac{\Gamma(a_n)}{\Gamma(a_0)} + (a_n - a_0) \, \psi(a_n) - (b_n - b_0) \, \frac{a_n}{b_n} \; .
\end{split}
\end{equation}

Using $x \sim \mathrm{Gam}(a, b) \Rightarrow \left\langle x \right\rangle = a/b$ again, it follows that

\begin{equation} \label{eq:GLM-NG-Com3}
\begin{split}
\mathrm{Com}(m) &= \frac{1}{2} \frac{a_n}{b_n} \left[ (\mu_0 - \mu_n)^T \Lambda_0 (\mu_0 - \mu_n) \right] + \frac{1}{2} \mathrm{tr}(\Lambda_0 \Lambda_n^{-1}) - \frac{1}{2} \log \frac{|\Lambda_0|}{|\Lambda_n|} - \frac{p}{2} \\
&+ a_0 \, \log \frac{b_n}{b_0} - \log \frac{\Gamma(a_n)}{\Gamma(a_0)} + (a_n - a_0) \, \psi(a_n) - (b_n - b_0) \, \frac{a_n}{b_n} \; .
\end{split}
\end{equation}

Rearranging and collecting the terms, we obtain the final result (Soch et al., 2016, eq.~C.4; Soch \& Allefeld, 2016a, eqs.~18/30):

\begin{equation} \label{eq:GLM-NG-Com4}
\begin{split}
\mathrm{Com}(m) = & + \frac{1}{2} \frac{a_n}{b_n} \left[(\mu_0-\mu_n)^T \Lambda_0 (\mu_0-\mu_n) - 2(b_n-b_0)\right] \\
&+ \frac{1}{2} \mathrm{tr}(\Lambda_0 \Lambda_n^{-1}) - \frac{1}{2} \log \frac{|\Lambda_0|}{|\Lambda_n|} - \frac{p}{2} \\
&+ a_0 \log \frac{b_n}{b_0} - \log \frac{\Gamma(a_n)}{\Gamma(a_0)} + (a_n - a_0) \psi(a_n) \; .
\end{split}
\end{equation}

\subsubsection{Log model evidence for the GLM-NG}

With the help of formulas for posterior hyperparameters (Soch et al., 2016, eqs.~A.9/10), one can indeed show that the difference of model accuracy and model complexity equals the log model evidence (Soch et al., 2016, eq.~B.9)

\begin{equation} \label{eq:GLM-NG-LME}
\begin{split}
\mathrm{LME}(m) = \frac{1}{2} & \log |P| - \frac{n}{2} \log (2 \pi)  + \frac{1}{2} \log |\Lambda_0| - \frac{1}{2} \log |\Lambda_n| \\ + & \log \Gamma(a_n) - \log \Gamma(a_0) + a_0 \log b_0 - a_n \log b_n \; .
\end{split}
\end{equation}

The calculation of individual accuracies and complexities (see eqs.~\ref{eq:GLM-NG-Acc4}/\ref{eq:GLM-NG-Com4}) is implemented in \verb|ME_GLM_NG_AnC|. Just like log model evidences, $\mathrm{Acc}(m)$ and $\mathrm{Com}(m)$ are calculated in a cross-validated fashion, so that cross-validated accuracy and complexity are given as sums of out-of-sample accuracies and complexities

\begin{equation} \label{eq:cvAnC}
\begin{split}
\mathrm{cvAcc}(m) &= \sum_{i=1}^{S} \mathrm{oosAcc}_i(m) \\
\mathrm{cvCom}(m) &= \sum_{i=1}^{S} \mathrm{oosCom}_i(m)
\end{split}
\end{equation}

and for each CV fold $i$, it holds that (see eq.~\ref{eq:LME-AnC})

\begin{equation} \label{eq:oosLME-oosAnC}
\mathrm{oosLME}_i(m) = \mathrm{oosAcc}_i(m) - \mathrm{oosCom}_i(m) \;.
\end{equation}

\subsection{Derivation and calculation of family evidences from cvLMEs}

The cvLME is an attempt to calculate the (logarithmized) model evidence $p(y|m)$ without prior information. Therefore, the (exponentiated) cvLME is taken as a substitute for the model evidence in all operations building on $p(y|m)$ such as Bayes factors, posterior probabilities, family evidences (Soch et al., 2016, eqs.~16-18) and group-level model selection (Soch et al., 2016, eqs.~D.1/2).

The relation between model evidences and family evidences is very simple and follows from the law of marginal probability:

\begin{equation} \label{eq:FE}
p(y|f) = \sum_{m \in f}^{} p(y|m) \, p(m|f) \; .
\end{equation}

If a uniform prior over models within each family is assumed, the family evidence is just the average of the model evidences

\begin{equation} \label{eq:FE-up}
p(y|f) = \frac{1}{M_f} \sum_{i=1}^{M_f} p(y|m_i)
\end{equation}

where $M_f$ is the number of models in family $f$. This sounds very simple at first glance, but the problem is that we usually cannot access model evidences $p(y|m)$ directly, but only deal with log model evidences $\log p(y|m)$. LMEs are used to avoid computational problems with very small model evidences that could not be stored in standard computers, e.g. $p(y|m) = 10^{-100} \Rightarrow \log p(y|m) \approx -230$. However, just exponentiating LMEs does not work, because they often fall below a specific underflow threshold $-u$, e.g. $u = 745$, so that all model evidences would be $0$.

The solution is to select the maximum LME within a family

\begin{equation} \label{eq:LME-max}
\mathrm{L}^{*}(f) = \max_{m \in f} \left[ \mathrm{LME}(m) \right]
\end{equation}

and define differences between LMEs and maximum LME as

\begin{equation} \label{eq:LME-diff}
\mathrm{L}'(m) = \mathrm{LME}(m) - \mathrm{L}^{*}(f) \; .
\end{equation}

\pagebreak
Then, the \textit{log family evidence} (LFE) can be written as

\begin{equation} \label{eq:LFE}
\mathrm{LFE}(f) = \log p(y|f) = \log \left[ \frac{1}{M_f} \sum_{i=1}^{M_f} \exp \left[ \mathrm{LME}(m_i) \right] \right]
\end{equation}

which can be further developed in the following way:

\begin{equation} \label{eq:LFE-LME}
\begin{split}
\mathrm{LFE}(f) &= \log \left[ \frac{1}{M_f} \sum_{i=1}^{M_f} \exp \left[ \mathrm{L}'(m_i) + \mathrm{L}^{*}(f) \right] \right] \\
&= \log \left[ \frac{1}{M_f} \exp \mathrm{L}^{*}(f) \sum_{i=1}^{M_f} \exp \mathrm{L}'(m_i) \right] \\
&= \mathrm{L}^{*}(f) + \log \sum_{i=1}^{M_f} \exp \mathrm{L}'(m_i) - \log M_f \; .
\end{split}
\end{equation}

In this way, only the differences $\mathrm{L}'(m_i)$ between LMEs and maximum LME must be exponentiated. If a difference is smaller than $-u$, contribution from the respective model $m_i$ will be automatically ignored -- and rightfully so, since it is much less evident than the best model in the family in this case.

If a non-uniform within-family prior distribution $p(m|f)$ is assumed, the above approximation (see eq.~\ref{eq:LFE-LME}) does not hold. In this case, we have to refomulate as follows:

\begin{equation} \label{eq:FE-nup}
\begin{split}
p(y|f) &= \sum_{m \in f}^{} p(y|m) \, p(m|f) \\
&= \frac{1}{M_f} \sum_{i=1}^{M_f} p(y|m_i) \, p(m_i|f) \, M_f \; .
\end{split}
\end{equation}

Then, the same procedure can be applied, just that LMEs need to be updated as

\begin{equation} \label{eq:LME-nup}
\mathrm{LME}'(m_i) = \mathrm{LME}(m_i) + \log p(m_i|f) + \log M_f \; ,
\end{equation}

such that their exponential corresponds to the product term in each summand:

\begin{equation} \label{eq:ME-add}
\exp \left[ \mathrm{LME}'(m_i) \right] = p(y|m_i) \, p(m_i|f) \, M_f \; .
\end{equation}

These procedures are implemented in \verb|MA_LFE_uniform| and \verb|MA_MF_LFE|.

\pagebreak
\subsection{Efficient estimation of random-effects BMS}

The cvBMS approach requires a voxel-wise implementation of \textit{random-effects Bayesian model selection} (RFX BMS), a population proportion model commonly used in \textit{dynamic causal modeling} (DCM) when seeking to perform group-level model selection. Although RFX BMS is implemented in SPM as \verb|spm_BMS|, we have re-implemented this technique in the MACS toolbox as \verb|ME_BMS_RFX_VB|, because the SPM routine is optimized for performing a single RFX BMS estimation (over a set of DCMs) as opposed to the mass-univariate estimation (over voxel-wise GLMs) required here.

Due to the \textit{Variational Bayesian} (VB) \textit{inversion scheme} for this hierarchical Bayesian model, there is no way to circumvent voxel-wise estimation of RFX BMS by multiple calls to \verb|ME_BMS_RFX_VB| -- in contrast to e.g. \verb|ME_GLM_NG| where a lot of computations are voxel-independent or can be performed jointly for all voxels. However, we have replaced loops over models and subjects in the SPM implementation by the respective vector or matrix computations in the MACS toolbox. In this way, we were able to speed up computation by a factor of 10 (Soch \& Allefeld, 2015, tab.~1). 

A second and more important change concerns an efficient calculation of exceedance probabilities following RFX BMS by using numerical integration over Gamma distributions instead of the established sampling from a Dirichlet distribution (see next section). This led to further speed improvements.

\subsection{Efficient calculation of EPs after RFX BMS}

The result of RFX BMS is a posterior distribution $p(r|y)$ over model frequencies $r$ which informs us about how probable all possible combinations of model frequencies are relative to each other. It typically concentrates most probability mass towards high values for the model which is most likely given the data.

To quantify this, one calculates \textit{exceedance probabilities} (EPs) as

\begin{equation} \label{eq:Dir-EP-def}
\varphi_j = p \left( \forall i \in \left\lbrace 1,\ldots,k | i \neq j \right\rbrace: \, r_j > r_i |\alpha \right) = p \left( \bigwedge_{i \neq j} r_j > r_i | \alpha \right) \; ,
\end{equation}

where $r_1, \ldots, r_k$ are model frequencies, $\alpha_1, \ldots, \alpha_k$ are the concentration parameters of the posterior Dirichlet distribution $p(r|y) = \mathrm{Dir}(r;\alpha)$ and $k$ is the number of models, so that $\varphi_j$ is the posterior probability of model $j$ being more frequent in the population than all the other models, given the sample drawn from the population.

If there are only two models, the Dirichlet distribution reduces to a Beta distribution, such that EPs can be calculated as (Soch \& Allefeld, 2016b, eq.~12)

\begin{equation} \label{eq:Dir2-EP}
\varphi_1 = 1 - \frac{\mathrm{B} \left( \frac{1}{2};\alpha_1,\alpha_2 \right)}{\mathrm{B}(\alpha_1,\alpha_2)} \quad \text{and} \quad \varphi_2 = 1 - \varphi_1
\end{equation}

where $\mathrm{B}(\alpha,\beta)$ is the beta function and $\mathrm{B}(x;\alpha,\beta)$ is the incomplete beta function. If the number of models is larger than two, exceedance probabilities cannot be calculated in this simple way and another approach has to be used. Here, we review the established method as well as our alternative proposal for estimating exceedance probabilities.

\subsubsection{Sampling from Dirichlet distribution}

Using the first method, exceedance probabilities are calculated via sampling from the respective distribution. Dirichlet random numbers can be generated by first drawing $q_1, \ldots, q_k$ from independent gamma distributions with shape parameters $\alpha_1, \ldots, \alpha_k$ and rate parameters $\beta_1 = \ldots = \beta_k$ and then dividing each $q_j$ by the sum of all $q_j$. This makes use of the relation (Soch \& Allefeld, 2016b, eq.~15)

\begin{equation} \label{eq:Gam-Dir-AB}
\begin{split}
& Y_1 \sim \mathrm{Gam}(\alpha_1,\beta), \, \ldots, \, Y_k \sim \mathrm{Gam}(\alpha_k,\beta), \, Y_s = \sum_{j=1}^k Y_j \\
\Rightarrow \; & X = (X_1, \ldots, X_k) = \left( \frac{Y_1}{Y_s}, \ldots, \frac{Y_k}{Y_s} \right) \sim \mathrm{Dir}(\alpha_1, \ldots, \alpha_k)
\end{split}
\end{equation}

where the probability density function of the gamma distribution is given by

\begin{equation} \label{eq:Gam-pdf}
\mathrm{Gam}(y; a, b) = \frac{{b}^{a}}{\Gamma(a)} \, y^{a-1} \, \exp[-b y] \quad \text{for} \quad y > 0 \; .
\end{equation}

Upon random number generation, exceedance probabilities can be estimated as

\begin{equation} \label{eq:Dir-EP1}
\varphi_j = \frac{1}{S} \sum_{n=1}^{S} \left[ \bigwedge_{i \neq j} r_j^{(n)} > r_i^{(n)} \right]
\end{equation}

where $[ \ldots ]$ refers to Iverson bracket notation, $S$ is the number of samples and $r_j^{(n)}$ corresponds to the $j$-th element from the $n$-th sample of $r$. Unfortunately, sampling is time-consuming and precise estimation of Dirichlet EPs can require up to $S = 10^6$ samples. We therefore propose another method relying on numerical integration.

\subsubsection{Integration over Gamma distributions}

Using this second method, exceedance probabilities are again calculated using theorem (\ref{eq:Gam-Dir-AB}). Therefore, consider

\begin{equation} \label{eq:Gam-Dir-A}
q_1 \sim \mathrm{Gam}(\alpha_1,1), \, \ldots, \, q_k \sim \mathrm{Gam}(\alpha_k,1), \, q_s = \sum_{j=1}^k q_j
\end{equation}

and the Dirichlet variate

\begin{equation} \label{eq:Gam-Dir-B}
r = (r_1, \ldots, r_k) = \left( \frac{q_1}{q_s}, \ldots, \frac{q_k}{q_s} \right) \sim \mathrm{Dir}(\alpha_1, \ldots, \alpha_k) \; .
\end{equation}

Obviously, it holds that

\begin{equation} \label{eq:Gam-Dir-eq}
r_j > r_i \; \Leftrightarrow \; q_j > q_i \quad \text{for} \quad i,j = 1, \ldots, k \quad \text{with} \quad i \neq j \; .
\end{equation}

Therefore, consider the probability that $q_j$ is larger than $q_i$, given $q_j$ is known. This probability is equal to the probability that $q_i$ is smaller than $q_j$, given $q_j$ is known

\begin{equation} \label{eq:Gam-EP0}
p(q_j > q_i|q_j) = p(q_i < q_j|q_j)
\end{equation}

which can be expressed in terms of the gamma cumulative distribution function as

\begin{equation} \label{eq:Gam-EP1}
p(q_i < q_j|q_j) = \int_0^{q_j} \mathrm{Gam}(q_i;\alpha_i,1) \, \mathrm{d}q_i = \frac{\gamma(\alpha_i,q_j)}{\Gamma(\alpha_i)}
\end{equation}

where $\Gamma(\alpha)$ is the gamma function and $\gamma(\alpha,x)$ is the lower incomplete gamma function. Since the gamma variates are independent of each other, these probabilties factorize:

\begin{equation} \label{eq:Gam-EP2}
p(\forall_{i \neq j} \left[ q_j > q_i \right]|q_j) = \prod_{i \neq j} p(q_j > q_i|q_j) = \prod_{i \neq j} \frac{\gamma(\alpha_i,q_j)}{\Gamma(\alpha_i)} \; .
\end{equation}

Although it can be easily calculated using implementations of the gamma function and the lower incomplete gamma function in numerical software packages, this probability is still dependent on $q_j$. In order to obtain the exceedance probability $\varphi_j$, $q_j$ has to be integrated out. From equations (\ref{eq:Dir-EP-def}) and (\ref{eq:Gam-Dir-eq}), it follows that

\begin{equation} \label{eq:Dir-EP2a}
\varphi_j = p(\forall_{i \neq j} \left[ r_j > r_i \right]) = p(\forall_{i \neq j} \left[ q_j > q_i \right]) \; .
\end{equation}

Using the law of marginal probability, we have

\begin{equation} \label{eq:Dir-EP2b}
\varphi_j = \int_0^\infty p(\forall_{i \neq j} \left[ q_j > q_i \right]|q_j) \, p(q_j) \, \mathrm{d}q_j \; .
\end{equation}

With (\ref{eq:Gam-EP2}) and (\ref{eq:Gam-Dir-A}), this becomes

\begin{equation} \label{eq:Dir-EP2c}
\varphi_j = \int_0^\infty \prod_{i \neq j} \left( p(q_j > q_i|q_j) \right) \, \mathrm{Gam}(q_j;\alpha_j,1) \, \mathrm{d}q_j \; .
\end{equation}

And with (\ref{eq:Gam-EP1}) and (\ref{eq:Gam-pdf}), it becomes

\begin{equation} \label{eq:Dir-EP2}
\varphi_j = \int_0^\infty \prod_{i \neq j} \left( \frac{\gamma(\alpha_i,q_j)}{\Gamma(\alpha_i)} \right) \, \frac{q_j^{\alpha_j-1} \exp[-q_j]}{\Gamma(\alpha_j)} \, \mathrm{d}q_j \; .
\end{equation}

In other words, the exceedance probability for each model amounts to an integral from zero to infinity where the first term in the integrand conforms to a product of gamma cumulative distribution functions and the second term is a gamma probability density function  (Soch \& Allefeld, 2016b, eq.~27).

This procedure has been implemented as MACS function \verb|MD_Dir_exc_prob| (Soch \& Allefeld, 2016b, sec.~2.4) which replaces the SPM version \verb|spm_dirichlet_exceedance| and like \verb|ME_BMS_RFX_VB| is being called from \verb|MS_BMS_group| (Soch \& Allefeld, 2018, fig.~1). Using model spaces of different size, we have shown that numerical integration speeds up computation of EPs by a factor of 7 to 10, even when only using $S = 10^5$ for the sampling approach (Soch \& Allefeld, 2016b, p.~9) which is below the number of samples $S = 10^6$ recommended by SPM.

\pagebreak
\section{Cross-validated Bayesian model averaging (cvBMA)} \label{sec:cvBMA}

\subsection{Averaging of first-level parameter estimates from SPM}

In traditional \textit{Bayesian model averaging} (BMA), the posterior distribution over model parameters $\theta$ is given by

\begin{equation} \label{eq:BMA}
p(\theta|y) = \sum_{i=1}^{M} p(\theta|y,m_i) \, p(m_i|y)
\end{equation}

where $p(m_i|y)$ is the $i$-th model's \textit{posterior probability} (PP) and $p(\theta|y,m_i)$ is the posterior distribution over $\theta$ given $m_i$.

In our case of model averaging across GLMs for fMRI, as we are only focusing on the regression coefficients $\beta$ and since we want to work with the parameter estimates provided by SPM, this reformulates to

\begin{equation} \label{eq:BMA-GLM}
\hat{\beta}_{\mathrm{BMA}} = \sum_{i=1}^{M} \hat{\beta}_i \cdot p(m_i|y)
\end{equation}

where $\hat{\beta}_i$ is the $i$-th model's point estimate for a given regression coefficient, usually obtained using \textit{restricted maximum likelihood} (ReML) and \textit{weighted least squares} (WLS) estimation in SPM.

Because we combine BMA with cross-validation across sessions, there are two possibilies of model averaging here. First, BMA could be performed session-wise using oosLMEs \textit{before} averaging the model-averaged parameter estimates across sessions to obtain one parameter estimate (\textit{session-wise} or \textit{out-of-sample BMA}):

\begin{equation} \label{eq:oosBMA}
\hat{\beta}_{\mathrm{oosBMA}} = \frac{1}{S} \sum_{j=1}^{S} \left( \sum_{i=1}^{M} \hat{\beta}_{ij} \cdot p(m_i|y_j) \right)
\end{equation}

where $p(m_i|y_j)$ is the PP of the $i$-th model calculated from the oosLME in the $j$-th session.\linebreak[4] Second, BMA could be performed across sessions using the cvLME \textit{after} averaging parameter estimates across sessions (\textit{session-wide} or \textit{cross-validated BMA}):

\begin{equation} \label{eq:cvBMA}
\hat{\beta}_{\mathrm{cvBMA}} = \sum_{i=1}^{M} \left( \frac{1}{S} \sum_{j=1}^{S} \hat{\beta}_{ij} \right) \cdot p(m_i|y)
\end{equation}

where $p(m_i|y_j)$ is the PP of the $i$-th model calculated from the cvLME across all sessions. Note that both formulas (\ref{eq:oosBMA}) and (\ref{eq:cvBMA}) can be rearranged into the form

\begin{equation} \label{eq:sessBMA}
\hat{\beta}_{\mathrm{BMA}} = \sum_{i=1}^{M} \frac{1}{S} \sum_{j=1}^{S} \left( \hat{\beta}_{ij} \cdot \mathrm{PP}(m_i) \right) \; .
\end{equation}

For two reasons, we have decided for the approach of cross-validated BMA: First, the average of \textit{maximum-likelihood} (ML) estimates across sessions is equivalent to the \textit{maximum-a-posteriori} (MAP) estimate when analyzing all data (Soch et al., 2017, eq.~A.5) which allows us to stay in the SPM workflow by building on its parameter estimates.

Second and more importantly, this second approach is also likely to be more precise due to the following reasoning: Imagine two models which differ by (around) 1 in their oosLMEs in five sessions, so that they differ by 5 in the cvLME across all sessions. When comparing two models and using a uniform prior over models, the \textit{log Bayes factor} (LBF) gives rise to PPs as (Soch \& Allefeld, 2018, eq.~20):

\begin{equation} \label{eq:PP-LBF}
p(m_1|y) = \frac{\exp(\mathrm{LBF}_{12})}{\exp(\mathrm{LBF}_{12})+1} = \frac{\exp(\mathrm{LME}_{1}-\mathrm{LME}_{2})}{\exp(\mathrm{LME}_{1}-\mathrm{LME}_{2})+1} \; .
\end{equation}

Consequently, the (average) posterior probability of the favored model will be around 0.73 when using the oosLMEs, but around 0.99 when using the cvLME. Therefore, by building on a posterior probability that is informed by more data, parameter estimates will on average get closer to the true values, given that the cvLME on average favors the true model to a stronger extent than oosLMEs do.

The procedure described here (see eq.~\ref{eq:cvBMA}) is implemented in \verb|MS_BMA_subject|. For single-session fMRI data where across-session averaging does not apply, just one parameter estimate for each regressor enters BMA with the cvLME.

\subsection{Efficient calculation of posterior probabilities for BMA}

In standard \textit{Bayesian model averaging} (BMA), posterior model probabilities are calculated using Bayes' theorem

\begin{equation} \label{eq:PP}
p(m_i|y) = \frac{p(y|m_i) \, p(m_i)}{\sum_{j=1}^M p(y|m_j) \, p(m_j)}
\end{equation}

where $p(y|m_i)$ is the $i$-th model evidence and $p(m_i)$ is the prior probability of the $i$-th model which -- assuming all models are equally likely \textit{a priori} -- is usually taken from a discrete uniform distribution $p(m) = 1/M$.

Again, simply exponentiating $\mathrm{cvLME}(m_i)$ to replace $p(y|m_i)$ causes problems, because cvLMEs are typically smaller than the underflow threshold. However, because posterior probabilities do not depend on absolute LME \textit{values}, but only on relative LME \textit{differences} or, equivalently, ME \textit{ratios}, the average across models can be removed from LMEs without changing the posterior probabilities in (\ref{eq:PP}). Therefore, the voxel-wise mean LME is subtracted from LMEs before they are exponentiated, multiplied with the prior and normalized across models according to (\ref{eq:PP}).

A voxel-wide version of this procedure is implemented in \verb|MS_BMA_subject| as follows:

\begin{verbatim}
    prior = 1/M * ones(M,1);
    LMEp  = LME - repmat(mean(LME,1),[M 1]);
    LMEp  = exp(LMEp) .* repmat(prior,[1 V]);
    post  = LMEp ./ repmat(sum(LMEp,1),[M 1]);
\end{verbatim}

Here, an $M \times V$ (models $\times$ voxels) matrix of LMEs is transformed into an $M \times V$ matrix of PPs which can be multiplied element-wise with an $M \times V$ matrix of $\hat{\beta}$ values for a given regressor to avoid voxel-wise computation and achieve efficient BMA.

\pagebreak
\section{Software} \label{sec:Software}

Originally, cvBMS and cvBMA were released as separate toolkits\footnote{URL: \url{https://github.com/JoramSoch/cvBMS}.}\textsuperscript{,}\footnote{URL: \url{https://github.com/JoramSoch/cvBMA}.} in December 2016 and March 2017. These methods had to be used in a command-line style when applying the respective operations to GLMs estimated in SPM.

With the introduction of MACS\footnote{URL: \url{https://github.com/JoramSoch/MACS}.} in May 2017, these toolkits have become obsolete. All operations in the MACS toolbox, including cvBMS and cvBMA, are completely accessible through a GUI and can be flexibly combined using the SPM batch editor. Extensive documentation of the MACS toolbox can be found in the the corresponding paper (Soch \& Allefeld, 2018). A toolbox manual can be obtained from the GitHub repository (Soch, 2018). The MACS toolbox is optimized for MATALB R2013b and SPM12, but also works with MATLAB R2007b and SPM8 or later.

\section{References} \label{sec:References}

\renewcommand{\section}[2]{}

\end{document}